Topic: Resilience and Climate Change

# An activity-based spatial-temporal community electricity vulnerability assessment framework

Chen Xia[1], Yuqing Hu[2], Jianli Chen[3]

1Department of Architectural Engineering, The Pennsylvania State University, University Park, PA 16802, United States of America; e-mail: cpx5037@psu.edu
2Department of Architectural Engineering, The Pennsylvania State University, University Park, PA 16802, United States of America; e-mail: yfh5204@psu.edu
3Department of Civil and Environmental Engineering, University of Utah, Salt Lake City, UT 84112, United States of America; email: jianli.chen@utah.edu

*Corresponding email:* yfh5204@psu.edu

## SUMMARY
The power system is among the most important critical infrastructures in urban cities and is getting increasingly essential in supporting people's daily activities. However, it is also susceptible to most natural disasters such as tsunamis, floods, or earthquakes. Electricity vulnerability, therefore, forms a crucial basis for community resilience. This paper aims to present an assessment framework of spatial-temporal electricity vulnerability to support the building of community resilience against power outages. The framework includes vulnerability indexes in terms of occupant demographics, occupant activity patterns, and urban building characteristics. To integrate factors in these aspects, we also proposed a process as activity simulation-mapping-evaluation-visualization to apply the framework and visualize results. This framework can help planners make an effective first-time response by identifying the most vulnerable areas when a massive power outage happens during natural disasters. It can also be integrated into community resilience analysis models and potentially contributes to effective disaster risk management

## INTRODUCTION
With electrification gaining ground in equipment, vehicles, and HVAC system, electricity is getting increasingly essential in supporting people's daily activities. According to the household site energy consumption survey conducted by US Energy Information Administration in 2015 [1], 47% of US household energy source relies on electricity. In the south part of the region, this percentage grows to as high as 69%. The electricity percentage is expected to continually increase with the trend of electrification [2]. However, the power system is susceptible to most natural disasters, either extreme heat or extreme cold can induce grid failures. For instance, cold waves and ice can break the lines, while heatwaves can lower the capacity of an electric line up to the point of overloading it [3]. This led to risks in the community that highly rely on electricity, as extreme weather and natural disaster often come with high electricity demand. Nationally, the US insurance industry has identified a USD 20–55 billion annual economic loss from power outages caused by flooding, hurricanes, and extreme temperature [4]. In the recent power outage caused by three severe winter storms sweeping across the United States in February 2021, more than 4.5 million homes and businesses were left without power, and nearly 700 people were estimated killed by the storm during the week with the worst power outages [14]. Since recent years have seen a trend of more natural disasters globally, the vulnerabilities of community power systems should be minimized to deal with various sources of disruption.

The definition of vulnerability in terms of climate change falls into two categories. The first refers to the potential damage caused to an electricity system when it is exposed to hazards [5]. Under this definition, research on power system vulnerability assessment is performed on the physical supply infrastructure system. The assessment indices include the anticipated loss of power source, broken degree of the power grid, the system's sensitivity to the impacts of the hazard, etc. [6]. Generally, complexity network methods are used to evaluate how vulnerability point collapse influences the whole power system performance [7]. Others focus on power system operation communication against electrical catastrophes based on decision theory. For example, Cheng et al [8] analyzed the vulnerability of a power grid using an optimization model based on game theory. In their analysis model, the grid operator and attacker are both actively adjusting their strategy to defend and bring down the power grid respectively, during which process vulnerable and critical power component was identified through optimization programming methods.

The other definition of vulnerability is more focused on the demand side, and measures vulnerability based on potential economic damage or human mortality and morbidity [9]. A common way is to quantify customer interruption cost (CIC) through a customer survey or blackout case study [10]. Indices of social vulnerability assessment framework related to unspecific events often focus on personal and household status on physical, economic, and social status [11]. Most of these factors are related to the socio-demographic feature of occupants and report a constant status of vulnerability. However, vulnerability changes with the varying exposure status of occupants to power outage risks. Studies have already revealed how household daily activity patterns influence the amount of electricity consumption, which directly influences power lines load, especially in extreme weather with high concentrated power demand [12]. Besides vulnerability caused by socio-demographic features identified in those electricity customer risk surveys, the activity people performed may also make them vulnerable to unexpected power outages. So, in addition to occupant socio-demographic characteristics, other influences like activity patterns should also be considered when evaluating how severe the effect that power outage may cause to the community is.

In this research, we propose a spatial-temporal electricity assessment framework that includes factors concerned with occupant activities, occupant demographics, and building characteristics to systematically assess a community's vulnerability on the demand side. To reach this goal, a model to simulate occupancy activities was built and mapped onto geographic information through linking a set of spatial and non-spatial databases. In addition to occupant and activity distribution information, building environments are further considered to assess the electricity vulnerability assessment. Finally, the vulnerability results are added to the geographic information and visualized to help get a better understanding of the potential impact of power outages on a community spatially and temporally.

**COMMUNITY ELECTRICITY VULNERABILITY ASSESSMENT FRAMEWORK**
Indexes in the community vulnerability assessment framework often include both general determinants and factors specific to events or disasters. For example, the quality of housing, as a special factor, will be a critical factor to impact a community's vulnerability to earthquakes or storms but is less likely to influence its vulnerability to drought [9]. In this study, both general and electricity-specific factors in the aspect of occupant socio-demographic characteristics, occupant activity, and building environment are identified through literature review. As Figure 1 shows, social demographic information in terms of physical, economic, educational, and social status are identified as general factors that determine the vulnerability of individuals and communities to a range of different hazards. These factors have been well studied and discussed in community vulnerability analysis research. For example, Spielman et al [13] proposed a set of Social Vulnerability Index (SoVI) and identified 28 demographic variables. They ranked these variables in terms of contributions to social vulnerability, among which the highest importance is age.

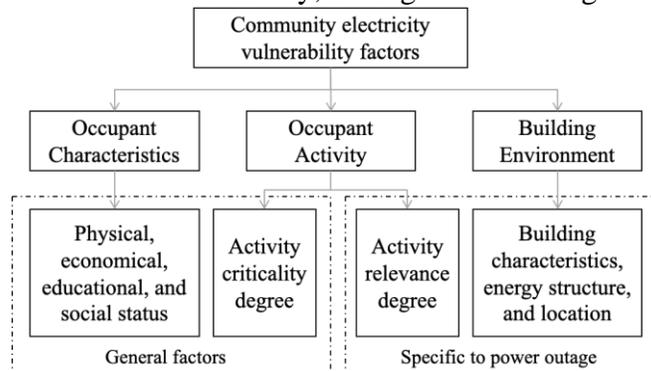

*Figure 1. community electricity vulnerability index structure*

Occupant activity assessment indexes could be divided as criticality and relevance. The criticality indexes are developed as general factors, where high criticality means the activity is essential for an occupant's daily life. Once it is interrupted, it will have a severe influence on the occupant (e.g., health-related activities). The relevance indexes are developed as electricity-specific factors based on the connection of electricity and specific activity. For example, typically, laundry activity has a higher relevance with electricity than general household chores. On some occasions, high relevance activity may have a low criticality degree such as watching TV or playing computer games, while some critical activity such as eating and sleeping does not rely so much on electricity in general conditions. Therefore, we integrate both indexes in activity-level electricity vulnerability assessment. Building environment indexes focus on specific factors concerning power outages. Besides direct factors such as energy structure, research reveals that building characteristics (e.g., age, size, construction, glazing) and building location could also influence household electricity consumption behaviours [15].

    To integrate factors in the three aspects, the process of conducting the assessment framework is proposed in Figure 2. The first step is to simulate time-activity trajectory, which is aimed to link information of occupant features, occupant activity, and related building types together at dynamic periods. Since buildings with the same type still vary in detailed environment characteristics, the next step is performed to map time-activity trajectories from building type to specific buildings through linking a set of spatial and non-spatial databases (eg. Zillow, PolicyMap, OpenStreetMap). Once this step is completed, these factors could be integrated, measured, and turned into indexes to evaluate the community electricity vulnerability through the developed assessment framework. Finally, the result could be visualized with spatial-temporal vulnerability information linked to a geographic information system. The following part will explain how each process is performed in detail.

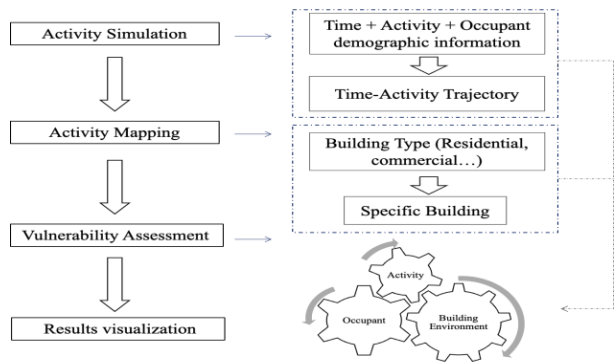

*Figure 2. Community electricity vulnerability assess process*

**TIME-ACTIVITY TRAJECTORY SIMULATION**

Two principal elements in the time-activity trajectory are time and activity. To simulate an activity schedule, we divide 24 hours of a day into 96 time periods of 15 minutes and estimate all activity probabilities during these periods. As for the activity category, there is no standard uniform classification of daily activities yet. The most accepted typology in travel behavior studies is proposed by Reichman, which divides activities into three categories as subsistence, maintenance, and discretionary (leisure) activities [16]. Kitamura et al [17] later advocated a simplified activity classification of two categories: mandatory (must-engaged activity) and discretionary (individual has the choice to be engaged). Other classifications like physiological needs, institutional demands, personal obligations, and personal preferences are also employed in research [18]. This study refers to original activity categories in the ATUS (American Time Use Survey) and standards in the literature and divides the activity into 8 classifications as follows: c01-essential health activity, c02-biological needs (e.g., eating, sleeping), c03-working, c04-education, c05-household management, c06-personal obligations (e.g., shopping, banking, childcare, etc), c07-personal preference (leisure activities), and c08-others (outside traveling activities).

Existing activity modeling methods could be categorized into mechanism-identified methods and data-driven methods. Rule-based methods focused on underlying mechanisms of how people make decisions on 24-hour time use among different activities. Data-driven models focus on extracting the activity patterns of a specific group of people through data mining techniques. For example, Jiang et al [19] used the principal component analysis (PCA) and K-means clustering algorithm to cluster several representative groups based on activity data of 30,000 individuals and then compared the social demographic differences in each cluster. Stating that traditional principal component analysis only represents frequent activities and disregard the infrequent ones, Liu et al [20] developed a novel process derived from Hidden Markov Models (pHMMs) to quantify the occurrence probabilities and sequence of all daily activities based on a large-scale activity-travel diary dataset.

Compared with models figuring out specific activity decision mechanisms, the data-driven model can quickly update community trajectory according to dynamic data accessed and can solve uncertainty underlying the human decision-making process. Therefore, in this study, we use the data-driven model and apply the Markov chain to simulate community trajectories. Markov chain has an attribute that the probability of observation in any state is only influenced by the preceding states and is often used to simulate the stochastic process. It has also been widely employed in activity simulation research [21]. For a discrete sequence of states, given the first-order Markov chain with an initial probability α and a transition matrix $\xi_{pq}$. the probability of transitions from state p to state q in sequence *i* could be calculated with the stationary transition matrix $\xi^i_{pq}$. Once we define an initial activity in the start time step (e.g., Sleeping at 3 am), we can get the probability of each activity in the following time step sequences.

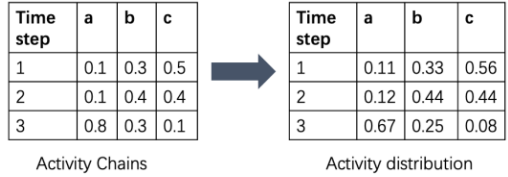

*Figure 3. Time-activity pattern transformation*

With the activity occurrence probability in each time step, we then integrate all activity probabilities into one period and get the broken distribution of community time-activity trajectory. For example, as Figure 3 shows if we have three activities A, B, C at time steps 1, 2, 3. Occurrence probability of activity A in each time step is 0.1,0.1,0.8, B is 0.3,0.3,0.4, C is 0.5,0.5,0; then we get the distribution of activities in time step 1 as A-0.11, B-0.33, C-0.56; time step 2 as A-0.11, B-0.33, C-0.56, time step 3 as A-0.11, B-0.33, C-0.56. For this study, the time-activity trajectory generated will be a 96*8-dimension matrix.

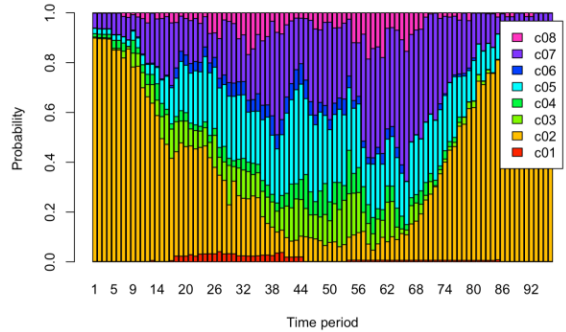

*Figure 4. Daily time-activity trajectory patterns*

(Note: c01- health emergency, c02-biological needs, c03- household management, c04-personal obligation, c05-working, c06-education, c07-personal preference, c08-others)

Figure 4 is an example that we run this model in a county of New York City. The x-axis denotes each 15 minutes time step, while the y-axis denotes the probability of each activity in that time step. An obvious trend in the figure 4 is that during the night, biological activity(c02) occupies the most possibilities. Though there are still some people awake for personal preference at night, there is a scarce probability for other activities. When time shifts to daytime, other activities begin to grow in possibility, especially the activity working, whose probabilities surges around time step 20 at 5 am. According to the probability distribution, we can then estimate the percentage of people performing corresponding activities.

**ACTIVITY-BUILDING MAPPING**

In the constructed time-activity trajectories, activities are tied to specific building types based on reported activity data. For example, housework activity is tied to residential buildings, while shopping is most likely linked to mercantile buildings. However, building environments still vary from building to building even with the same type. This step is to map time-activity trajectories from building type to specific buildings. One method we propose is to deploy different standards to map. Figure 5 shows the mapping framework. For residential buildings, we can assign people in each building according to the number of bedrooms and vacancy rate once we know the total number of people in this type of building in a specific period. Gross floor area and work density are similar standards for business buildings. Number and capacity can be used as standards for some building types such as mercantile, public service, and assembly buildings, as capacity determines the maximum number of people the specific building could hold, and the number of these buildings determines the maximum capacity for a specific area. The education building is a little different from other types of buildings. Besides features of the building itself like the type of school, community demographic information is also used to help identify people in different schools at different periods. Since activity trajectory will only report location as education building, the distribution of people in detail school type (eg, middle or high school) relies on the demographic structure.

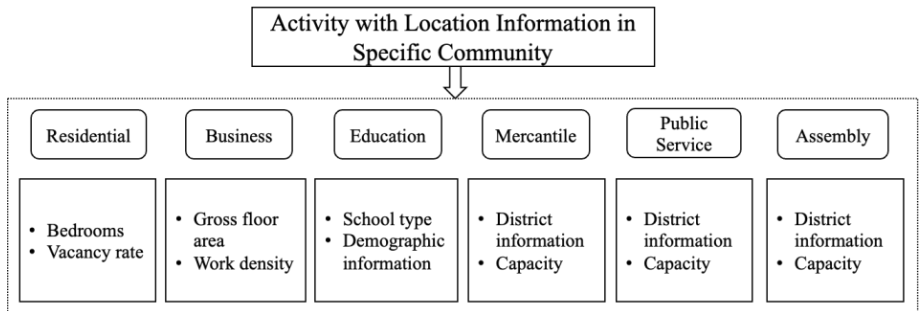

*Figure 5. Activity-building mapping framework*

Besides developing standards to perform the mapping, the development of modern technology of Geographic Information Systems (GIS) also throws some light on this process. People's time-location paths can be constructed through data collected with handheld GPS (Global Positioning System) units, GPS-enabled smartphone tracking applications, or A-GPS (assisted GPS) devices [22]. Compared to manually mapping activity to a specific building, digital devices that are either embedded in vehicles or carried by people in smartphones can report geographic location densely and automatically. Though it mostly only reports information about time and location, as Figure 6 shows, a connection between the time-location path and time-activity trajectory can be built based on time to construct the time-activity-building trajectory.

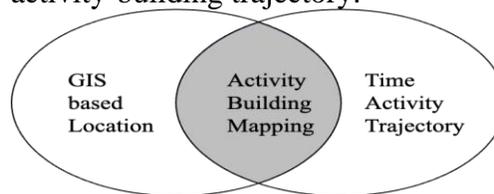

*Figure 6. Mapping with GIS aided methods*

**VULNERABILITY ASSESSMENT AND VISUALIZATION**

With indexes of occupant demographic, occupant activity, and building environment in the integrated framework, a Vulnerability Rating Index (VRI) is then established to measure the community electricity vulnerability. As proposed by Xu et al [23] in their construction of heat vulnerability index, to make sure the index is cohesive within aspects, rather than using absolute values, variable inside each aspect is measured and integrated into an overall ranking between 1 to 5 as low, medium-low, medium, medium-high, and high. Then each aspect will be assigned a weight according to a literature review and experts survey, which also allows users to adjust based on different needs in policy developments or emergency response planning. Finally, the overall rate could be calculated by integrating the weighted rankings of the three aspects. As defined in Equation 1, let $p_i$ a ranking of the $i$ aspect of the three index dimensions, $q_i$ a weight of this aspect contributing to the overall vulnerability, then the overall electricity vulnerability rate of the specific area of a community is calculated as:

$$V = \sum_i p_i \, x \, q_i, \quad (i = 1,2,3) \tag{1}$$

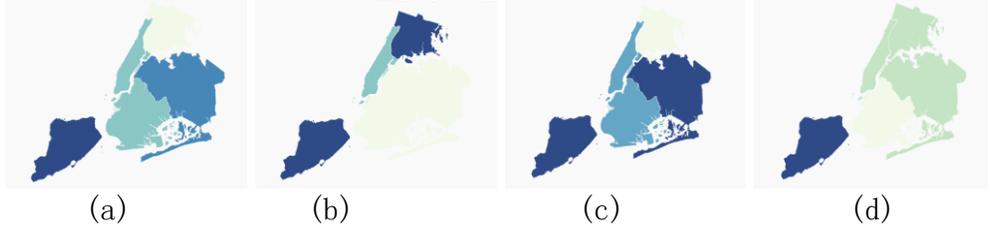

(a)         (b)         (c)         (d)

*Figure 7. Potential visualization results*

To visualize vulnerability degree in different areas of a specific community, we take a reference to Shach-Pinsly's study on security visualization of urban areas [24]. A separate GIS layer is created for vulnerability ranking results for each aspect, pixel displaying the transformed vulnerability degree of all measurements results inside each layer, where lighter color means lower vulnerability. To get the final electricity vulnerability degree, the pixel in each layer is then summed with weights and displayed in another separate map. Figure 7 shows potential visualization results: Figure 7(a) is the distribution of occupant demographic vulnerability. This is a general factor, deeper color reveals that people in this area are more vulnerable to any disaster rather than merely power outage-related events. Figure 7(b) is the occupant activity vulnerability distribution, which will be the major changing factor influenced by time. Figure 7(c) is the built environment vulnerability specific to a power outage. Figure 7(d) is the final community electricity vulnerability degree visualization results. In this example, for a specific period, with time changes by hour or day, visualization in Figure 7(b) in the map will change accordingly, making Figure 7(d) a spatial-temporal map of community electricity vulnerability. If time lasts long, building environments vulnerability in Figure 7(c) will also be updated.

**CONCLUSION**

In this article, we proposed an assessment framework of spatial-temporal urban electricity vulnerability to support the building of community resilience against power outages. The framework includes indexes in terms of occupant demographics, occupant activity patterns, and urban building characteristics. Indices of each aspect consider both the general vulnerability against unexpected events/disasters in the community and specific factors concerned with the power outage. To integrate all three aspects, we also proposed the process to apply this framework. A model to simulate occupancy activities in real-time was first built and then mapped onto geographic information through linking a set of spatial and non-spatial databases. With occupant and activity distribution information, building environments are further considered to assess the electricity vulnerability assessment. Finally, the vulnerability results are added to the geographic information and visualized to help get a better understanding of the potential impact of power outages on a community spatially and temporarily. This gives a hint to the power system manager to make better responses in the face of power outage events and may also assist the local government in the planning of infrastructure and resources to reduce residents' vulnerability to power outage events. In this framework, we only proposed a qualitative framework, more work will be done in the future on fulfilling this framework into an applicable quantified model

# REFERENCES

[1] U.S. Energy Information Administration, "Residential Energy Consumption Survey (RECS) - Household Energy Insecurity 2015"
[2] U.S. Energy Information Administration, " The International Energy Outlook (EIO) 2019"
[3] A. Abedi, L. Gaudard, and F. Romerio, "Review of major approaches to analyze vulnerability in power system," Reliability Engineering & System Safety, vol. 183, pp. 153–172, 2019.
[4] A. Korbatov, J. Price-Madison, Y. Wang and Y. Xu, "The risks of climate and natural disaster related disruption to the electric grid," Johns Hopkins University, 2016.
[5] R. Jones, and R. Boer. "Assessing current climate risks adaptation policy framework: a guide for policies to facilitate adaptation to climate change." UNDP, in review, see http://www. undp. org/cc/apf-outline. Htm, 2003.
[6] A. M. A. Haidar, A. Mohamed, en A. Hussain, "Vulnerability assessment of power system using various vulnerability indices", in 2006 4th Student Conference on Research and Development, Shah Alam, Malaysia, 2006.
[7] B. Wang, D. H. You, en X. G. Yin, "Research on vulnerability assessment system of complicated power system", Adv. Mat. Res., vol 204–210, bll 622–626, Feb 2011.
[8] M. X. Cheng, M. Crow, en Q. Ye, "A game theory approach to vulnerability analysis: Integrating power flows with topological analysis", Int. j. electr. power energy syst., vol 82, bll 29–36, Nov 2016.
[9] N. Brooks, "Vulnerability, risk and adaptation: A conceptual framework", Tyndall Centre for climate change research working paper, vol 38, bll 1–16, 2003.
[10] Q. Yan, T. Dokic, en M. Kezunovic, "GIS-based risk assessment for electric power consumers under severe weather conditions", in 2016 18th Mediterranean Electrotechnical Conference (MELECON), Lemesos, Cyprus, 2016.
[11] S. Fuchs en T. Thaler, Vulnerability and resilience to natural hazards. Cambridge University Press, 2018.
[12] S. Singh en A. Yassine, "Mining energy consumption behavior patterns for households in smart grid", IEEE Trans. Emerg. Top. Comput., vol 7, no 3, bll 404–419, Jul 2019.
[13] S. E. Spielman et al., "Evaluating social vulnerability indicators: criteria and their application to the Social Vulnerability Index", Nat. Hazards (Dordr.), vol 100, no 1, bll 417–436, Jan 2020.
[14] P. Aldhous, S. M. Lee, en Z. Hirji, "The Texas winter storm and power outages killed hundreds more people than the state says", BuzzFeed News, 2021.
[15] L. Giusti en M. Almoosawi, "Impact of building characteristics and occupants' behaviour on the electricity consumption of households in Abu Dhabi (UAE)", Energy Build., vol 151, bll 534–547, Sep 2017.
[16] S. Reichman, "Travel adjustments and life styles: a behavioral approach", in Behavioral Travel Demand Models, Lexington, Massachusetts, 1976.
[17] R. Kitamura, C. Chen, R. M. Pendyala, en R. Narayanan, "Micro-simulation of daily activity-travel patterns for travel demand forecasting", Transportation, vol 27, no 1, bll 25–51, 2000.
[18] B. Vilhelmson, "Daily mobility and the use of time for different activities. The case of Sweden", GeoJournal, vol 48, bll 177–185, 1999.
[19] S. Jiang, J. Ferreira, en M. C. González, "Clustering daily patterns of human activities in the city", Data Min. Knowl. Discov., vol 25, no 3, bll 478–510, Nov 2012.
[20] F. Liu, D. Janssens, J. Cui, G. Wets, en M. Cools, "Characterizing activity sequences using profile Hidden Markov Models", Expert Syst. Appl., vol 42, no 13, bll 5705–5722, Aug 2015.
[21] Y. Zhou, Q. Yuan, C. Yang, en Y. Wang, "Who you are determines how you travel: Clustering human activity patterns with a Markov-chain-based mixture model", Travel Behav. Soc., vol 24, bll 102–112, Jul 2021.
[22] F. Qi en F. Du, "Trajectory data analyses for pedestrian space-time activity study", J. Vis. Exp., no 72, bl e50130, Feb 2013.
[23] Y. Xu, T. Hong, W. Zhang, Z. Zeng, en M. Wei, Heat Vulnerability Index Development and Mapping. 2021.
[24] D. Shach-Pinsly, "Measuring security in the built environment: Evaluating urban vulnerability. in a human-scale urban form", *Landsc. Urban Plan.*, vol 191, no 103412, bl 103412, Nov 2019.